\documentclass[11pt,oneside,letterpaper]{article}
\usepackage{amssymb}
\usepackage{amsmath,bm}
\usepackage[dvips]{graphicx}
\usepackage{setspace}
\usepackage{fancyhdr}
\usepackage[dvipsnames]{xcolor}
\usepackage{ifpdf}
\usepackage{graphicx}
\usepackage{cite}
\usepackage{rotating}
\usepackage{comment}
\usepackage[margin=2.5cm]{geometry}
\usepackage[colorlinks=true, linkcolor  = blue,urlcolor=blue,citecolor=violet]{hyperref}
\usepackage{soul}
\usepackage{relsize}
\usepackage{tikz}
\usepackage{tikzlings}
\usetikzlibrary{trees}
\usetikzlibrary{decorations.pathmorphing}
\usetikzlibrary{decorations.markings}
\usetikzlibrary{decorations.markings}
\usepackage{float}
\usepackage{empheq}
\usepackage{bbold}
\usepackage{etoolbox}
\patchcmd{\thebibliography}{\section*{\refname}}{}{}{}

\newcommand{\bi}{\begin{itemize}}
\newcommand{\ei}{\end{itemize}}
\newcommand{\bea}{\begin{eqnarray}}
\newcommand{\eea}{\end{eqnarray}}
\newcommand{\be}{\begin{equation}}
\newcommand{\ee}{\end{equation}}

\newcommand{\dd}{\mathrm{d}}

\numberwithin{equation}{section}
\allowdisplaybreaks  

\begin{document}



\onehalfspacing

\begin{center}

~
\vskip4mm
{{\huge {
\quad The two-sphere partition function in two-dimensional quantum gravity at fixed area
 }
  }}
\vskip5mm

\vskip2mm

\vskip10mm


Beatrix M\"uhlmann \\ 

\end{center}
\vskip4mm
\begin{center}
{
\footnotesize
{Institute for Theoretical Physics and $\Delta$ Institute for Theoretical Physics, University of Amsterdam, \\Science Park 904, 1098 XH Amsterdam, The Netherlands\\
}}
\end{center}
\begin{center}
{\textsf{\footnotesize{
b.muhlmann@uva.nl}} } 
\end{center}
\vskip5mm

\vspace{4mm}
 
\vspace*{0.6cm}

\vspace*{1.5cm}
\begin{abstract}
We discuss two-dimensional quantum gravity coupled to conformal matter and fixed area in a semiclassical large and negative matter central charge limit. In this setup the gravity theory---otherwise highly fluctuating---admits a round two-sphere saddle. We discuss the two-sphere partition function up to two-loop order from the path integral perspective. This amounts to studying Feynman diagrams incorporating the fixed area constraint on the round two-sphere. In particular we find that all ultraviolet divergences cancel to this order. We compare our results with the two-sphere partition function obtained from the DOZZ formula.   
\noindent

\end{abstract}


\newpage

\tableofcontents

\section{Introduction}\label{sec:intro}
Two-dimensional quantum gravity differs from higher dimensional theories of gravity. In two dimensions the Einstein-Hilbert action is topological. The theory is over-constrained and has no locally propagating degrees of freedom. In fact the Einstein-Hilbert action is proportional to the Euler characteristic of the two-dimensional manifold it is integrated over. The path integral first sums over all geometries of a fixed genus and is then summing over all genera. In two dimensions we can furthermore exploit the two diffeomorphisms to write the metric in terms of a fixed fiducial part $\tilde{g}_{ij}$ and a single degree of freedom appearing in the form of the Weyl factor $\varphi(x)$. It has been postulated that in the Weyl gauge a theory of two-dimensional quantum gravity is given by a path integral over the Weyl factor weighted by the Liouville action \cite{Polyakov:1981rd, Distler:1988jt, David:1988hj} 
\begin{equation}
S_L[\varphi]= \frac{1}{4\pi}\int \dd^2 x\sqrt{\tilde{g}}\left(\tilde{g}^{ij}\partial_i \varphi\partial_j\varphi+ Q\tilde{R}\varphi+ 4\pi \Lambda\, e^{2b\varphi}\right)~.
\end{equation}
Here $\tilde{R}$ is the Ricci scalar of the fixed fiducial metric, $\Lambda \geq 0$ is the cosmological constant. Liouville theory is a two-dimensional CFT  \cite{Seiberg:1990eb,Zamolodchikov:2005fy,  Zamolodchikov:1995aa, Dorn:1994xn, Teschner:2001rv} and the parameters $b$ and $Q$ are related to the Liouville central charge through
\begin{equation}
c_L= 1+6Q^2~,\quad Q= b+b^{-1}~.
\end{equation}
In addition to a theory of pure two-dimensional quantum gravity one can also consider the addition of matter fields. Here we focus on a specific matter theory given by the series $\mathcal{M}_{2,2m-1}$ of minimal models  \cite{Polyakov:1984yq, Friedan}, forming themselves a conformal field theory with known central charge $c_{\text{m}}$ and operator content. In particular to avoid the appearance of a conformal anomaly we obtain the constraint
\begin{equation}
c_L+ c_{\text{m}}+c_{\text{gh}}=0~,
\end{equation}
where $c_{\text{gh}}=-26$ is the central charge of the $\mathfrak{b}\mathfrak{c}$-ghost theory, introduced upon restricting to Weyl gauge. \newline\newline
In general two-dimensional quantum gravity constitutes a highly fluctuating object. For genus zero we can tame these fluctuations in two different ways: 
\begin{itemize}
\item[] A) \quad For $c_\text{m}\rightarrow -\infty$  and fixed area we find a round sphere saddle \cite{Zamolodvarphikov:1982vx}
\item[] B) \quad For $c_\text{m}\rightarrow \infty$ the theory itself admits a round sphere saddle \cite{timelike, Polchinski:1989fn}
\end{itemize}
In either of these cases we obtain a round two-sphere saddle. In other words, we started with a a priori highly fluctuating metric of a genus zero surface and collapse it on a round two-sphere geometry. The fluctuations in either case are suppressed at order $\mathcal{O}(|c_\text{m}|^{-1})$. In this paper we focus on case A), whereas case B) was discussed in \cite{timelike}. \newline
The main motivation to study case A stems from the conjectured duality between two-dimensional quantum gravity coupled to the non-unitary minimal model $\mathcal{M}_{2,2m-1}$ with $c_{\text{m}}<0$ and a specific class of matrix integrals, known as multicritical matrix integrals \cite{Kazakov:1989bc,Staudacher:1989fy}. Multicritical matrix integrals might provide a microscopic picture of the gravitational path integral. Furthermore the two-sphere is the Euclidean realisation of two-dimensional de Sitter space \cite{dS_musings}. Very little is known about de Sitter space at the quantum level. In particular, because of the accelerated expansion and observer in a de Sitter spacetime is surrouded by a cosmological horizon. Conjecturally a finite entropy---whose microscopic origin is not understood---is assigned to this horizon \cite{Gibbons:1976ue, Gibbons:1977mu}, defined via the Euclidean path integral on compact surfaces. 

This paper is part of a series of papers \cite{Anninos:2020ccj,Anninos:2020geh,db} trying to relate techniques of matrix integrals and two-dimensional quantum gravity to study the Euclidean path integral of spacetimes with positive cosmological constant.
\newline\newline
The paper is structured as follows. In the first section \ref{sec:2d} we introduce the theory of two-dimensional quantum gravity and discuss the gauge fixing. In particular we explain how to gauge fix the residual symmetry group on the sphere, left also upon fixing the Weyl gauge. In section three \ref{sec:semi} we introduce and discuss the genus zero fixed area path integral semiclassically, i.e. in the limit of large negative matter central charge. We provide a semiclassical expansion up to two-loop contributions. This in particular involves the computations of Feynman diagrams on the two-sphere with an additional constraint -- the fixed area. In the last section \ref{sec:DOZZ} we compare to the DOZZ formula. Furthermore we compare the two-sphere partition functions obtained in case A) and case B). We end with some remarks relating the results obtained in this paper to multicritical matrix integrals. 

\section{Two-dimensional quantum gravity}\label{sec:2d}

The theory we will focus on consists of a two-dimensional matter CFT of central charge $c_{\text{m}}$ coupled to two-dimensional quantum gravity. Additionally we will fix the physical area $\int_{\Sigma_h} \dd x^2 \sqrt{g}= 4\pi\upsilon$, where $\upsilon$ is a real parameter and $\Sigma_h$ denotes a compact genus $h$ surface. The Euclidean path-integral of interest is given by \cite{Zamolodvarphikov:1982vx,Seiberg:1990eb}
\begin{equation}\label{zgrav0}
\mathcal{Z}_{\text{grav}}[\upsilon] = \sum_{h=0}^\infty e^{\vartheta \chi_h}\int [\mathcal{D} g_{ij} ] \, e^{- \Lambda \int_{\Sigma_h} \dd^2 x \sqrt{g} } \times Z^{(h)}_{\text{CFT}}[g_{ij}]\times \delta \left(\int_{\Sigma_h} \dd x^2 \sqrt{g}- 4\pi \upsilon\right)~,
\end{equation}
where we take $\Lambda >0$, $\chi_h$ is the Euler character of $\Sigma_h$, and $Z^{(h)}_{\text{CFT}}[g_{ij}]$ is the matter CFT partition function.  The positivity of $\Lambda$ suppresses large area configurations.
Although the pure gravity theory has no classical solutions in two-dimensions due to the topological nature of the Einstein term, upon fixing the area and coupling to a matter theory with $c_{\text{m}}\rightarrow -\infty$ the effective gravitational action including the contribution from $Z^{(h)}_{\text{CFT}}[g_{ij}]$ does \cite{Zamolodvarphikov:1982vx}. 

On a genus zero surface in complex coordinates the two-dimensional metric in the Weyl gauge can be expressed as
\begin{equation}\label{Weylcomplex}
ds^2 = e^{2b\varphi(z,\bar{z})}  \tilde{g}_{z\bar{z}} \, \text{d} z \, \text{d} \bar{z}~, \quad\quad z \in \mathbb{C}~.
\end{equation}
We have included the parameter $b$ for future convenience. Our choice of fiducial metric takes the Fubini-Study form
\begin{equation}\label{FSmetric}
d\tilde{s}^2 = \frac{4 \upsilon \text{d} {z} \text{d} \bar{z} }{\left(1+ z \bar{z}\right)^2}~.
\end{equation}
It will also be convenient to work with spherical coordinates
\begin{equation}
z = e^{i \phi} \, \tan \frac{\theta}{2}~, \quad\quad \bar{z} = e^{-i \phi} \, \tan \frac{\theta}{2}~,
\end{equation}
such that the fiducial metric is given by 
\begin{equation}\label{spherical}
d\tilde{s}^2 = \upsilon\left( \dd\theta^2 + \sin^2\theta \dd\phi^2 \right)~,
\end{equation}
with $\theta \in (0,\pi)$ and $\phi \sim \phi+2\pi$, capturing the two-sphere with area $4\pi \upsilon$. In order to properly fix the Weyl gauge (\ref{Weylcomplex}) we must introduce Fadeev-Popov $\mathfrak{b}\mathfrak{c}$-ghost fields. Upon integrating out the matter and $\mathfrak{b}\mathfrak{c}$-ghost fields, our resulting gravitational path integral on a genus zero surface is given by 
\begin{equation}\label{g0Z}
\mathcal{Z}_{\text{grav}}^{(0)}[\upsilon] =  e^{2\vartheta} \times \frac{\mathcal{A}}{\text{vol}_{PSL(2,\mathbb{C})}} \times  \upsilon^{(c_{\text{m}}-26)/6} \times \upsilon^{-1} \int [\mathcal{D}\varphi] \,  e^{-S_L[\varphi]} \times \delta \left(\int_{S^2} \dd \Omega\,e^{2b\varphi}-4\pi\right)~,
\end{equation}
where $\mathcal{A}$ captures contributions from the matter and CFT partition function at a sphere with reference area normalised to one.
According to the hypothesis of Distler-Kawai \cite{Distler:1988jt} and David \cite{David:1988hj}, this action is given by the Liouville action \cite{Polyakov:1981rd}
\begin{equation}\label{SL}
S_{L}[\varphi] = \frac{1}{4\pi} \int \dd^2x\sqrt{\tilde{g}} \left( \tilde{g}^{ij} \partial_i \varphi \partial_j \varphi +  Q \tilde{R} \varphi + 4\pi \Lambda e^{2b \varphi}  \right)~,
\end{equation}
with $[\mathcal{D}\varphi]$ being the standard flat measure on the space of fields $\varphi$. Moreover, $Q = b+1/b$ and $c_L = 1+6 Q^2$. Consistency of the theory, viewed as a theory of gravity coupled to conformal matter, requires $c_L - 26 + c_{\text{m}} = 0$. In terms of the matter central charge, 
\begin{equation}\label{Qb}
Q = \sqrt{\frac{25-c_{\text{m}}}{6}}~, \quad\quad b =  \frac{\sqrt{25-c_{\text{m}}}-\sqrt{1-c_{\text{m}}}}{2 \sqrt{6}}~,
\end{equation} 
where the positive root for $Q$ is a choice we can make given the redundancy $Q \to - Q~$, $b\to -b$, and $\varphi \to - \varphi$. 

\subsection{Residual gauge symmetries \& further gauge fixing.}\label{resgauge}

On a genus zero surface (\ref{Weylcomplex}) does not fully fix the gauge. For instance, the transformation
\begin{equation}\label{redundant}
2b\varphi(z,\bar{z}) \to 2b\varphi(z,\bar{z})  - \sigma(z,\bar{z}) ~, \quad\quad \tilde{g}_{ij}(z,\bar{z})  \to e^{\sigma(z,\bar{z}) } \tilde{g}_{ij}(z,\bar{z})~,
\end{equation}
is a redundancy of the parametrisation. As such, the resulting theory must be invariant under the above redundancy. Locally, conformal Killing vectors of $\tilde{g}_{ij}$ are given by holomorphic maps
\begin{equation}\label{holomorphic}
z \to f(z)~, \quad \bar{z} \to \overline{f(z)}~.
\end{equation}
The above maps do not affect the form of the Weyl gauge (\ref{Weylcomplex}) since they can be reabsorbed in a shift of $\varphi(z,\bar{z})$. Of the space of holomorphic maps, we must select the subset of (\ref{holomorphic}) which are normalisable under the norm on the space of diffeomorphisms $\omega^i(z,\bar{z})$ \cite{Polyakov:1981rd, Zamolodvarphikov:1982vx} 
\begin{equation}
ds_\omega^2 = \int  \text{d} {z} \text{d} \bar{z}  \sqrt{g} g_{ij} \omega^i \omega^j~.
\end{equation}
This leaves a $PSL(2,\mathbb{C})$ subgroup of normalisable residual diffeomorphisms. 
It is relatively straightforward to check that in the semiclassical limit the transformation
\begin{equation}\label{phiT}
\varphi(z,\bar{z}) \to \varphi(f(z),\overline{f(z)}) + \frac{Q}{2} \log f'(z) + \frac{Q}{2} \log \overline{f'(z)} + Q \left(\Omega(f(z),\overline{f(z)} ) - \Omega(z,\bar{z}) \right)~,
\end{equation}
leaves $S_{L}$ invariant.\footnote{Beyond this limit the classical Liouville action is no longer invariant, rather it is the quantum theory that must be invariant under (\ref{phiT}).}
Given the invariance of  $S_{L}$, and assuming it persists at the quantum level, the path-integral over the Liouville field $\varphi$ will produce a term proportional to the volume of $PSL(2,\mathbb{C})$. Thus, $\text{vol}_{PSL(2,\mathbb{C})}$ appears in both the numerator and denominator of the gravitational path integral (\ref{g0Z}) in the Weyl gauge. 

To fix this we must resort to further gauge-fixing. We can expand the Liouville field in a complete basis of real spherical harmonics
\begin{equation}\label{phiexp}
\varphi(\Omega) = \sum_{l,m} \varphi_{lm} Y_{lm}(\Omega)~,
\end{equation} 
with $\Omega$ a point on the round metric on $S^2$ with area $4\pi$, i.e. (\ref{spherical}) with $\upsilon=1$;  $\varphi_{lm}$ and $Y_{lm}(\Omega)$ are real valued.
 They obey the orthogonality relation
\begin{equation}\label{ortho}
\int \dd\Omega Y_{lm}(\Omega)Y_{l' m'}(\Omega) = \delta_{l l'} \delta_{m m'}~.
\end{equation}
Our conventions for spherical harmonics are given in appendix \ref{Yapp}. Under an isometric map $\Omega \to \Omega'(\Omega)$ the $Y_{lm}(\Omega)$ map to linear combinations of $Y_{lm}(\Omega)$ with the same $l$. 
{This is because for each $l$, $Y_{lm}(\Omega)$ with $m\in [-l,l]$ furnish irreducible representations of $SO(3)$.} As a gauge-fixing condition we follow \cite{Distler:1988jt} and impose that $\varphi_{1m} = 0$ with $m=\{-1,0,1\}$. $SO(3)$ invariance fixes the form of the above to be independent of the $\varphi_{1m}$ and ensures that $\Delta_{\text{FP}}$ takes the following structure
\begin{multline}\label{FPexact}
\Delta_{\text{FP}} = a_0 Q^3 + a_1 Q\sum_{m=-2}^2 \varphi^2_{2m} + a_2 \Big(\varphi_{2,0}^3+\frac{3}{2}\varphi_{2,0}(\varphi_{2,1}^2+\varphi_{2,-1}^2)+\frac{3}{2}\sqrt{3}\varphi_{2,2}(\varphi_{2,1}^2-\varphi_{2,-1}^2)\cr
+3\sqrt{3}\varphi_{2,1}\varphi_{2,-1}\varphi_{2,-2}-3\varphi_{2,0}(\varphi_{2,-2}^2+\varphi_{2,2}^2)\Big)~.
\end{multline}
where
\begin{equation}\label{a0a1a2}
a_0\equiv -\frac{16}{3\sqrt{3}}\pi^{3/2}~,\quad a_1\equiv \frac{12}{5}\sqrt{3\pi}~,\quad a_2\equiv \frac{12}{5}\sqrt{\frac{3}{5}}~.
\end{equation}
For further details we refer to \cite{timelike}.
At this stage, all but the $SO(3)$ isometry group of the original two-dimensional diffeomorphisms has been gauge fixed. Since this is a compact group, we can just divide out its volume explicitly. 
\section{Semiclassical saddle $\&$ small fluctuations}\label{sec:semi}
Imposing the area constraint the Liouville path integral on a genus zero surface is 
\begin{equation}\label{g0Z}
\mathcal{Z}_{L}[\upsilon] =  \frac{1}{\text{vol}_{PSL(2,\mathbb{C})}} \times \upsilon^{-1}e^{-4\pi \Lambda \upsilon} \int [\mathcal{D}\varphi] \,  e^{-S_L[\varphi,\Lambda=0]} \times \int_{\mathbb{R}}\frac{\dd \alpha}{2\pi}\, e^{i\alpha\,\left(\int_{S^2}\dd \Omega e^{2b\varphi}- 4\pi\right)}~, 
\end{equation}
where we introduced a Lagrange multiplier to fix the area.  The equations of motion of $\varphi$ and $\alpha$ are 
\begin{equation}\label{eom}
2\tilde{\nabla}^2\varphi= \frac{2}{\upsilon}Q- \frac{8}{\upsilon}\pi i b \alpha\,e^{2b\varphi}~,\quad \int_{S^2} \dd\Omega\,e^{2b\varphi}=4\pi~,
\end{equation}
where $\tilde{\nabla}^2$ is the Laplacian on the fiducial metric $\tilde{g}_{ij}$. The equations of motion (\ref{eom}) allow for the saddle 
\begin{equation}\label{saddle}
\varphi_*= 0~,\quad \alpha_*= -i\frac{Q}{4\pi  b}~,
\end{equation}
leading to a round two-sphere saddle for the physical metric (\ref{Weylcomplex}).
Due to the $PSL(2,\mathbb{C})$ invariance of the Liouville action, in addition to the above solution there is a continuous family of solutions related to (\ref{saddle}) by $PSL(2,\mathbb{C})$ transformations \cite{Zamolodvarphikov:1982vx}. Upon fixing the gauge of the residual gauge symmetries as discussed in section \ref{resgauge}, we collapse the continuous solution space down to the constant saddle. The saddle point then leads to 
\begin{equation}\label{Zsaddle}
\mathcal{Z}_{\mathrm{saddle}}[\upsilon]= \frac{1}{\mathrm{vol}_{SO(3)}}\times \upsilon^{-1} e^{-4\pi \Lambda \upsilon}~.
\end{equation}
We now recall that the semiclassical limit corresponds to $b\rightarrow 0$.
\subsection{One-loop contribution}
Expanding the Liouville field and the Lagrange multiplier around the semiclassical saddles $\varphi= \varphi_*+ \delta\varphi$, $\alpha= \alpha_*+ \delta\alpha$ (\ref{saddle}) we find 
\begin{equation}\label{Zpert}
\mathcal{Z}_{\mathrm{pert}}^{(2)}[b] =  \int [\mathcal{D}\delta\varphi] \times  \Delta_{\text{FP}}\times \prod_{m=\{-1,0,1\}} \delta(\delta\varphi_{1m}) \times \int_{\mathbb{R}}\frac{\mathrm{d}\delta\alpha}{2\pi}\, e^{-S^{(2)}_{\text{pert}}[\delta\varphi,\delta\alpha]}~,
\end{equation}
where
\begin{equation}\label{Spert}
S_{\mathrm{pert}}^{(2)}[\delta\varphi,\delta\alpha]
\equiv \frac{1}{4\pi} \int \dd x^2\sqrt{\tilde{g}}\left(\tilde{g}^{ij}\partial_i\delta \varphi\partial_j\delta \varphi - \frac{2}{\upsilon}(1+b^2)\delta\varphi^2 -4\pi \times \frac{2}{\upsilon}i b \delta\alpha\delta\varphi\right)~.
\end{equation}
Notice that the small correction in the mass is of the same order as the coefficient of the quartic interaction. In particular the mass term is a negative integer to leading order \cite{Bros:2010wa}. We now rescale $\delta\alpha\rightarrow \delta\alpha/b$ such that all quadratic pieces are of order one and expand $\delta \varphi$ in a complete and orthonormal basis of real spherical harmonics, as in (\ref{phiexp}). In terms of these, the measure of path integration is taken to be
\begin{equation}\label{measure}
[\mathcal{D}\delta\varphi] = \prod_{l,m} \left(\frac{\upsilon\Lambda_{\mathrm{uv}}}{\pi}\right)^{\frac{1}{2}}{\text{d}  \delta \varphi_{lm}}~,
\end{equation}
such that
\begin{equation}
1 = \int [\mathcal{D}\delta \varphi] e^{- \Lambda_{\mathrm{uv}} \int \dd^2 x \sqrt{\tilde{g}} \, \delta\varphi(x)^2}~.
\end{equation}
To render the measure local with respect to $\tilde{g}_{ij}$ we need to introduce an ultraviolet scale $\Lambda_{\text{uv}}$ with units of inverse area. The eigenvalues and degeneracies of the spherical Laplacian $-\tilde{\nabla}^2$ on (\ref{spherical}) are
\begin{equation}\label{eigenvalues}
\lambda_l = \frac{1}{\upsilon} \, l(l+1)~, \quad\quad d_l = 2l+1~, \quad\quad l = 0,1,\ldots~.
\end{equation}
At the Gaussian order, including the leading contribution stemming from the Fadeev-Popov determinant as well as the $1/b$ from the rescaling of $\delta\alpha$ we finally obtain 
\begin{equation}\label{eq:Z_oneloop}
\mathcal{Z}_{\mathrm{pert}}^{(2)}[b] =  \frac{a_0 Q^3}{b\upsilon}\left(\frac{\upsilon \Lambda_{\text{uv}}}{16\pi^2}\right)^{\frac{1}{2}}\left(\frac{\upsilon\Lambda_{\text{uv}}}{\pi}\right)^{\frac{3}{2}}\left(\frac{4\pi \upsilon\Lambda_{\text{uv}}}{6-2bQ-\frac{4\pi}{Q^2} \frac{a_1}{a_0}}\right)^{\frac{5}{2}}\prod_{l\geq 3}\left(\frac{4\pi \upsilon \Lambda_{\text{uv}}}{l(l+1)-2bQ}\right)^{l+\frac{1}{2}}~,
\end{equation}
where $a_1/a_0= -27/20\pi$ (\ref{a0a1a2}). In the above expression we strip out the $l=0, 1$ and $l=2$ modes. The $l=1$ modes are removed by our gauge fixing choice, whereas the $l=2$ modes are affected by the Fadeev-Popov determinant. From (\ref{Spert}) we infer that the $l=0$ modes couple to the Lagrange multiplier $\delta\alpha$. We have 
\begin{equation}
\int_{\mathbb{R}}\frac{\dd\delta\alpha}{2\pi}\, e^{2 i\delta\alpha\,\int \dd\Omega\,\varphi(\Omega)} = \int_{\mathbb{R}}\frac{\dd\delta\alpha}{2\pi}\, e^{4\sqrt{\pi} i\delta\alpha\,\delta\varphi_{00}}=\frac{1}{4\sqrt{\pi}}\times\delta\left(\delta\varphi_{00}\right)~.
\end{equation}
This explains the second term in (\ref{eq:Z_oneloop}).
The only subtle term is the infinite product which we regularise using a heat kernel analysis. We obtain (see e.g. \cite{loooongSphere})
\begin{multline}\label{infiniteProduct}
-\frac{1}{2} \sum_{l=3}^\infty ({2l+1} ) \log \left( \frac{l(l+1) -2bQ}{4\pi\Lambda_{\mathrm{uv}} \upsilon} \right) =-\frac{107+12\nu^2}{12}\log\left(\frac{2e^{-\gamma_E}}{\varepsilon}\right)+\frac{2}{\varepsilon^2}+\nu^2+\frac{3}{2}\log b^2\cr
+\frac{5}{2}\log(2-b^2)+ \frac{1}{2}\log(1+b^2)+ \frac{9}{2}\log 2+\left(\frac{1}{2}-\Delta_+\right)\zeta'(0,\Delta_+)\cr
+ \left(\frac{1}{2}-\Delta_-\right)\zeta'(0,\Delta_-)
+\zeta'(-1,\Delta_+)+\zeta'(-1,\Delta_-)~,
\end{multline}
where
\begin{equation}
\Delta_\pm\equiv \frac{1}{2}\pm i\nu~,\quad \nu\equiv i\sqrt{\frac{1}{4}+2bQ}~,
\end{equation}
and $\zeta(a,z)$ denotes the Hurwitz-$\zeta$ function.
 The heat kernel regularization amounts to
\begin{equation}\label{bessel}
-\frac{1}{2} \log \frac{{\color{black}\tilde{\lambda}}}{4\pi \upsilon\Lambda_{\mathrm{uv}}} = \int_0^\infty \frac{\dd\tau}{2\tau} e^{-\frac{\varepsilon^2}{4\tau}-{\color{black}\tilde{\lambda}}\tau} =  K_0\left(\sqrt{{\color{black}\tilde{\lambda}} } \varepsilon \right) \approx -\frac{1}{2}\log \frac{\varepsilon^2\, e^{2\gamma_E}{{\color{black}\tilde{\lambda}}}}{4}~,
\end{equation}
where $\varepsilon$ is a small parameter given by $\varepsilon = {e^{-\gamma_E}}/{\sqrt{\pi \upsilon \Lambda_{\mathrm{uv}}}}$, and $\gamma_E$ denotes the Euler-Mascheroni constant. The $1/\varepsilon^2$ divergence is local with respect to $\tilde{g}_{ij}$ and can be absorbed into an appropriate local counterterm built from $\tilde{g}_{ij}$. The parameter $\tilde{\lambda}= \upsilon\lambda \in
 \mathbb{R}^+$ denotes an eigenvalue.
Applying to (\ref{infiniteProduct}) the relations \cite{Adamchik}\footnote{These identities are to be understood as yielding a real valued analytic expression at small $b$.} 
\begin{equation}
\zeta'(0,z)= \text{log}\Gamma(z)-\frac{1}{2}\log(2\pi)~,\quad \zeta'(-1,z)= \zeta'(-1) -\log G(z+1)+z \, \text{log}\Gamma(z)~,
\end{equation} 
with $G(z)$ the Barnes $G$ function, $\zeta'(-1)=1/12-\log A$ with $A$ denoting Glaisher's constant, and $\text{log}\Gamma(z)$ the logGamma function we obtain the small $b$ expansion of the Hurwitz-$\zeta$ functions in (\ref{infiniteProduct})
\begin{eqnarray}
-\frac{1}{2} \sum_{l=3}^\infty ({2l+1} ) \log \left( \frac{l(l+1) -2bQ}{4\pi\Lambda_{\mathrm{uv}} \upsilon} \right)&\approx& -\frac{25}{12}+ \frac{20}{3}\gamma_E+ \frac{1}{2}\log 6+ \log 96-2 \log A-\frac{20}{3}\log\frac{2}{\varepsilon}\cr
&+& b^2 \left(2\log\frac{2}{\varepsilon}- \frac{37}{12}\right)+ \mathcal{O}(b^4)~.
\end{eqnarray}
Combining (\ref{Zsaddle}) with (\ref{eq:Z_oneloop}) and using (\ref{infiniteProduct}) we obtain the small $b$ expansion of the fixed-area Liouville sphere partition function
\begin{equation}\label{eq:1loop_semi}
\mathcal{Z}_{L}[\upsilon]= \mathrm{const}\times \upsilon^{\frac{1}{6}+b^2}\,\Lambda_{\text{uv}}^{\frac{7}{6}+b^2}e^{-4\pi \Lambda\upsilon}\,\left(\frac{1}{b^4}+\left(\frac{7}{6}+ \frac{5a_1}{2a_0}\pi +2\gamma_E+ \log(4\pi)\right)\frac{1}{b^2}+ \ldots \right)~,
\end{equation}
where the constant is given by 
\begin{equation}\label{const}
\mathrm{const}\equiv \frac{a_0}{\mathrm{vol}_{SO(3)}}\, \frac{3\sqrt{3}}{8\times 2^{1/6}A^2\pi^{10/3}}\, e^{-25/12}~.
\end{equation}
In particular we highlight the $\upsilon$ dependency of (\ref{eq:1loop_semi}). Whereas Liouville theory is a two-dimensional CFT, which would force the sphere partition function to scale as $\upsilon^{c_L/6}$  \cite{Polyakov:1981rd,Zamolodchikov:2001dz}, where $c_L= 6/b^2+13+6b^2$ (\ref{Qb}), the area constraint (\ref{g0Z}) destroys conformality, leading to a different $\upsilon$ dependency of the fixed area Liouville sphere partition function. 
\subsection{Two-loop contributions} 
Upon expanding in spherical harmonics it becomes clear that in the perturbative action (\ref{Spert}) the $l=0$ mode couples to the fluctuation $\delta\alpha$.  We diagonalise the action to avoid this coupling at the Gaussian level
\begin{equation}\label{shift}
 \delta\varphi_{00}\rightarrow \delta\varphi_{00}- 4\pi^{3/2}\,i \delta\alpha~.
 \end{equation}
We thus obtain
\begin{equation}\label{interactions2}
\int [\mathcal{D}'\delta\varphi]\,\int_{\mathbb{R}}\frac{\dd\delta\alpha}{2\pi b}\,  e^{-\hat{S}^{(2)}_{\text{pert}}[\delta\varphi]}\,e^{+\frac{1}{2\pi}\delta\varphi_{00}^2}\,e^{+{8\pi^2}\delta\alpha^2}\times \,\left(e^{\frac{1}{4\pi}\int\dd\Omega\,\left(\frac{4}{3}b\delta\varphi^3+\frac{2}{3}b^2\delta\varphi^4\ldots \right)}\times e^{i\delta\alpha\int \dd \Omega \left(2b \delta\varphi^2+ \ldots\right)}\right) ~,
\end{equation}
 where $\hat{S}^{(2)}_{\text{pert}}[\delta\varphi]$ follows from (\ref{Spert}) upon splitting off the $l=0$ and $\delta\alpha$ modes, which we treat separately; the $l=1$ modes we removed through the gauge-fixing. Furthermore for the term in brackets we need not forget to shift the $l=0$ modes using (\ref{shift}). The primed measure indicates that we removed the three $l=1$ modes. At the order at hand we have $Qb=1$.

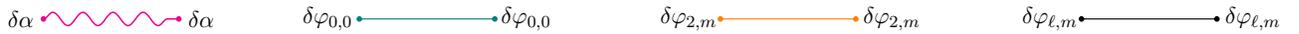
\begin{figure}[H]
\begin{center}
\begin{tikzpicture}[scale=.6]
\draw[decorate,decoration={snake,segment length=.4cm}, line width= 0.2mm,magenta] (-3,0) --(0,0);
\draw [fill,magenta] (0,0) circle [radius=0.05];
\draw [fill,magenta] (-3,0) circle [radius=0.05];
\node[scale=.8] at (0.5,0)   {$\delta\alpha$};
\node[scale=.8] at (-3.5,0)   {$\delta\alpha$};

\draw[line width=0.15mm,teal] (4,0) --(7,0);
\draw [fill,teal] (4,0) circle [radius=0.05];
\draw [fill,teal] (7,0) circle [radius=0.05];
\node[scale=.8] at (3.3,0)   {$\delta\varphi_{0,0}$};
\node[scale=.8] at (7.7,0)   {$\delta\varphi_{0,0}$};

\draw[line width=0.15mm,orange] (12,0) --(15,0);
\draw [fill,orange] (12,0) circle [radius=0.05];
\draw [fill,orange] (15,0) circle [radius=0.05];
\node[scale=.8] at (11.3,0)   {$\delta\varphi_{2,m}$};
\node[scale=.8] at (15.8,0)   {$\delta\varphi_{2,m}$};

\draw[line width=0.15mm] (20,0) --(23,0);
\draw [fill] (20,0) circle [radius=0.05];
\draw [fill] (23,0) circle [radius=0.05];
\node[scale=.8] at (19.3,0)   {$\delta\varphi_{\ell,m}$};
\node[scale=.8] at (23.8,0)   {$\delta\varphi_{\ell,m}$};
\end{tikzpicture}
\end{center}
\caption{Propagators of the small fluctuations.}
\label{fig:diagramsValpha}
\end{figure}

Whereas the $l=1$ modes which lead to an almost zero modes in (\ref{Spert}) we removed through our gauge fixing procedure, this action still contains one subtlety. Both the $l=0$ as well as the $\delta\alpha$ fluctuation lead to Gaussian unsuppressed terms in (\ref{Zpert}). A priori it seems that we have four choices 
\begin{align}\label{Wick4}
1)\quad \delta\varphi_{00} \rightarrow \pm i \delta\varphi_{00}~,\quad \delta\alpha \rightarrow \mp i\delta\alpha~,\cr
2) \quad \delta\varphi_{00} \rightarrow \pm i \delta\varphi_{00}~,\quad \delta\alpha \rightarrow \pm i\delta\alpha~.
\end{align}
In case 1) the resulting Jacobians cancel each other, whereas in case 2) we obtain an overall minus sign. In performing the Gaussian integrals, however we observe that only the two choices with $\delta\alpha\rightarrow -i\delta\alpha$ lead to cancellations of the UV divergences, which as we show in appendix \ref{app:unsuppressed}\footnote{I would like to thank Dio Anninos for pointing this approach out to me.} by proposing an ansatz which does not make use of Wick rotations, is the correct result.  In terms of Wick rotations we are thus left with
\begin{align}\label{Wickrotation}
a)\quad  &\delta\varphi_{00} \rightarrow  i \delta\varphi_{00}~,\quad \delta\alpha \rightarrow -i\delta\alpha~,\cr
b)\quad  &\delta\varphi_{00} \rightarrow  -i \delta\varphi_{00}~,\quad \delta\alpha \rightarrow -i\delta\alpha~.
\end{align}
In particular we observe that it does not lead to any sign ambiguity. This seems to suggest that even though both options (\ref{Wickrotation}) lead to the cancellations of the ultraviolet divergences, only opposite Wick rotations, i.e. case a) where the mutual determinants cancel, is consistent. We will discuss further evidence that in the next section \ref{sec:DOZZ} when we compare to the DOZZ formula. 
\newline
\newline
\textbf{Propagators.} Since we have to distinguish the $l=0,1$ and $l=2$ modes from the $l\geq 3$ modes we will work in momentum space. Using the expansion (\ref{phiexp}) and the orthonormality (\ref{ortho}) we have
\begin{equation}\label{propagator}
 \frac{1}{\int [\mathcal{D}'\delta\varphi] e^{-\hat{S}_{\mathrm{pert}}^{(2)}[\delta\varphi]}\, e^{-\frac{1}{2\pi}\delta\varphi_{00}^2}} \times \int [\mathcal{D}'\delta\varphi] e^{-\hat{S}_{\mathrm{pert}}^{(2)}[\delta\varphi]}\, e^{-\frac{1}{2\pi}\delta\varphi_{00}^2} \delta\varphi_{l,m}\delta\varphi_{l',m'} =\frac{2\pi}{A_l}\delta_{l,l'}\delta_{m,m'}~,
\end{equation}
where we defined for $l\neq 1$, $A_l\equiv (l(l+1)-2-2b^2)$.
Next we have the propagator for $\delta\alpha$. After Wick rotating we obtain in the semiclassical limit
\begin{equation}\label{eq:alpha_prop}
 \frac{1}{\int \dd\delta\alpha\, e^{-{8\pi^2}\delta\alpha^2}}\times \int \dd\delta\alpha \, e^{-{8\pi^2}\delta\alpha^2}\delta\alpha^2 =\frac{1}{16\pi^2}~.
\end{equation}
We are led to four distinct types of propagators as shown in the figure below (fig.\ref{fig:diagramsValpha})
\noindent
As a final remark, before delving into calculating the Gaussian integrals we note that upon performing the shift (\ref{shift}) as well as the Wick rotation (\ref{Wickrotation}) all Gaussian integrals involving powers of the $l=0$ mode vanish
\begin{align}\label{eq:holomorphic}
\int \dd \delta\varphi_{00}\dd \delta\alpha \, e^{-\frac{1}{2\pi}\delta\varphi_{00}^2}\,e^{-{8\pi^2}\delta\alpha^2}\left(\pm i \delta\varphi_{00} -{4\pi^{3/2}}\delta\alpha\right)^n=0~,\quad n> 0~.
\end{align}
Similarly we find 
\begin{equation}\label{eq:holomorphic2}
\int \dd \delta\varphi_{00}\dd \delta\alpha \, e^{-\frac{1}{2\pi}\delta\varphi_{00}^2}\,e^{-{8\pi^2}\delta\alpha^2}\left(\pm i \delta\varphi_{00} -{4\pi^{3/2}}\delta\alpha\right)^n (\delta\alpha)^m= (-1)^n2n!\left(\frac{1}{4\pi^{3/2}}\right)^{n+1}\delta_{m,n}~,\quad n,m > 0~.
\end{equation}
Now we turn to performing the Gaussian integrals in (\ref{interactions2}) after expanding to order $\mathcal{O}(b^2)$.\newline\newline
\textbf{Gaussian integrals.} We are left with the two possibilities (\ref{Wickrotation}). We explain one of these two cases, namely
\begin{equation}\label{eq:Wick_app}
\delta\varphi_{00}\rightarrow i\delta\varphi_{00}~,\quad \delta\alpha\rightarrow - i\delta\alpha~.
\end{equation}
as the intermediate steps, in particular the cancellation of UV divergences we will encounter, is unaffected by our choice. We will re-instore the sign ambiguity at the end. Combining (\ref{interactions2}) with (\ref{eq:Wick_app}) we have in the leading small $b$ expansion
 \begin{align}\label{interactions2}
\mathcal{Z}_{\text{pert}}[b]&=\int [\mathcal{D}'\delta\varphi]\,\int_{\mathbb{R}}\frac{\dd\delta\alpha}{2\pi b}\, e^{-\hat{S}_{\text{pert}}^{(2)}[\delta\varphi]}\,e^{-\frac{1}{2\pi}\delta\varphi_{00}^2}\,e^{-{8\pi^2}\delta\alpha^2}\cr
&\times \,\left(e^{\frac{1}{4\pi}\int\dd\Omega\,\left(\frac{4}{3}b\delta\varphi(\Omega)^3+\frac{2}{3}b^2\delta\varphi(\Omega)^4\ldots \right)}\times e^{\delta\alpha\int \dd \Omega \left(2b \delta\varphi(\Omega)^2+ \ldots\right)}\right) ~,
\end{align}
where for the term in brackets we need not forget to shift $\delta\varphi_{00}\rightarrow  i\delta\varphi_{00}- {4\pi^{3/2}}\, \delta\alpha$. Expanding the exponential we get up to order $\mathcal{O}(b^2)$
\allowdisplaybreaks
\begin{align}\label{eq:Zpertb2}
\mathcal{Z}_{\text{pert}}[b]&=\int [\mathcal{D}'\delta\varphi]\,\int_{\mathbb{R}}\frac{\dd\delta\alpha}{2\pi b}\, e^{-\hat{S}_{\text{pert}}^{(2)}[\delta\varphi]}\,e^{-\frac{1}{2\pi}\delta\varphi_{00}^2}\,e^{-{8\pi^2}\delta\alpha^2}\cr
&\times \left(1+ \frac{1}{6\pi}b^2\int \dd\Omega \delta\varphi(\Omega)^4+ \frac{1}{18\pi^2}b^2 \int\dd\Omega\dd\Omega' \delta\varphi(\Omega)^3\delta\varphi(\Omega')^3+ \ldots \right)\cr
&\times \left(1+ \frac{4}{3}b^2 \delta \alpha \int \dd\Omega\delta\varphi(\Omega)^3+2b^2 \delta\alpha^2 \int\dd\Omega\dd\Omega' \delta\varphi(\Omega)^2\delta\varphi(\Omega')^2+ \ldots \right)~\cr
&\times \left(1+ \frac{2}{3\pi}b^2 \delta\alpha\int \dd\Omega \dd\Omega' \delta\varphi(\Omega)^3\delta\varphi(\Omega')^2\right)~.
\end{align}
From the above we see that at order $\mathcal{O}(b^2)$ we encounter cubic and quartic vertices. Furthermore the fluctuation of the Lagrange multiplier $\alpha$ introduced to fix the area can interact with the $l=0$ mode of the field fluctuation $\delta\varphi$. 
We now treat the five terms in (\ref{eq:Zpertb2}) independently. We start with the three terms mixing the $\delta\alpha$ and $\delta\varphi$ fluctuations. 
\newline\newline
\textbf{Bobbles.} To obtain a non-vanishing Gaussian integral the fields need to appear in even powers. The only non-vanishing possibility to combine $\delta\alpha$ with a cubic $\delta\varphi$ fluctuation therefore implies that one of the spherical harmonics is a zero-mode. Using (\ref{eq:holomorphic}-\ref{eq:holomorphic2}) as well as the shift (\ref{shift}) combined with (\ref{eq:Wick_app}) we find the following logarithmically divergent contribution
\begin{align}\label{b1}
\bigcirc_1\equiv \frac{4}{3}b^2 \Big\langle\delta \alpha \int \dd\Omega\delta\varphi(\Omega)^3 \Big\rangle&= \frac{4}{3}b^2\times 3\times \frac{1}{2\sqrt{\pi}}\times (-4\pi^{3/2})\times\frac{1}{16\pi^2}\times 2\pi\times \sum_{l\geq 2}\frac{(2l+1)}{A_l}\cr
&=- b^2\sum_{l\geq 2}\frac{(2l+1)}{A_l}~.
\end{align}
We now explain all of these factors. The $4b^2/3$ stems from expanding the exponential, the $3$ is a symmetry factor, the $1/(2\sqrt{\pi})$ is the $l=0$ spherical harmonic we need in order to obtain an even power in $\delta\alpha$. For the $l=0$ mode we then use the shift (\ref{shift}) combined with (\ref{eq:Wick_app}) leading to ($-4\pi^{3/2}$). Finally we use the propagators (\ref{eq:alpha_prop}) and (\ref{propagator}), leading to the factors $1/(16\pi^2)$ and $2\pi$. All the other terms in (\ref{b1}) cancel upon using (\ref{eq:holomorphic}-\ref{eq:holomorphic2}).
\newline\newline
\textbf{Cacti and bobbles.} Following this logic we obtain 
\begin{align}\label{b2}
\bigcirc\!\!\bigcirc_1 +\bigcirc\!\!\bigcirc &+ \bigcirc_2=2b^2  \Big\langle\delta\alpha^2 \int\dd\Omega\dd\Omega' \delta\varphi(\Omega)^2\delta\varphi(\Omega')^2 \Big\rangle &\cr
&= \frac{b^2}{2} \sum_{l\neq l'\geq 2}\frac{(2l+1)(2l'+1)}{A_l A_{l'}}+\frac{b^2}{2} \sum_{l\geq 2}\frac{(2l+1)(2l+3)}{A_l^2}+ b^2\sum_{l\geq 2}\frac{(2l+1)}{A_l}~.
\end{align}
Of these the first one is double-logarithmically divergent, whereas the last one diverges logarithmically. We also obtain a finite contribution. \newline\newline
\textbf{Cacti.}
Finally we have 
\begin{align}\label{c2}
\bigcirc\!\!\bigcirc_2 +\bigcirc\!\!\bigcirc&=\frac{2}{3\pi}b^2 \Big\langle \delta\alpha\int \dd\Omega \dd\Omega' \delta\varphi(\Omega)^3\delta\varphi(\Omega')^2 \Big\rangle \cr
&=-b^2 \sum_{l\neq l'\geq 2}\frac{(2l+1)(2l'+1)}{A_l A_{l'}}- b^2\sum_{l\geq 2} \frac{(2l+1)(2l+3)}{A_l^2}~.
\end{align}
Of these, the first one is double-logarithmically divergent, whereas the second term yields a finite contribution. In particular it very important that the divergences hear appear with a minus sign. This rules out two cases in (\ref{Wick4}).
\newline\newline
We now turn to the $\delta\alpha$ independent terms. We have two distinct types of iteractions at order $\mathcal{O}(b^2)$: 
\begin{align}\label{alpha_ind}
\bigcirc\!\!\bigcirc_3+\mathlarger{\mathlarger{\ominus}}+\bigcirc\!\!-\!\!\bigcirc=&\frac{1}{6\pi}b^2\Big\langle\int \dd\Omega \delta\varphi(\Omega)^4\Big\rangle+  \frac{1}{18\pi^2}b^2 \Big\langle\int\dd\Omega\dd\Omega' \delta\varphi(\Omega)^3\delta\varphi(\Omega')^3\Big\rangle~.
\end{align}
We now discuss these three contributions.
 \newline\newline
\textbf{Cacti.} Performing Gaussian integrals for the quartic term in (\ref{alpha_ind}) we obtain what we will denote as cactus diagrams
\begin{equation}\label{cactus1}
\bigcirc\!\!\bigcirc_3 \equiv   \frac{1}{2}b^2\sum_{l,l'\geq 2}\frac{(2l+1)(2l'+1)}{A_l A_{l'}}~.
\end{equation}
This contribution diverges like the square of the logarithm.\newline\newline
\textbf{Melons.} The sextic term in (\ref{alpha_ind})  a priori has two contributions. Melons and double-tadpoles. However using (\ref{eq:holomorphic}) it is straight forward to see that double-tadpoles vanish. The melonic contribution yields
\begin{equation}\label{melon}
\mathlarger{\mathlarger{\ominus}} \equiv  \frac{1}{18\pi^2} \, c_1\, (2\pi)^3b^2\sum_{\bold{l}\geq \textbf{2},\bold{m}}  \, \frac{W_{\bold{l},\bold{m}} W_{\bold{l},\bold{m}}}{A_{l_1} A_{l_2} A_{l_3}}~. 
\end{equation}
where $\bold{l} = (l_1,l_2,l_3)$, $\bold{m} = (m_1,m_2,m_3)$, and
\begin{equation}\label{eq:real3j}
W_{\bold{l},\bold{m}} \equiv \int\dd\Omega \, Y_{l_1 m_1}(\Omega)Y_{l_2 m_2}(\Omega)Y_{l_3 m_3}(\Omega)~,
\end{equation}
is a type of Wigner 3-j symbol for real spherical harmonics; $c_1=6$ is a combinatorial factor.
To evaluate (\ref{eq:real3j}) we replace real spherical harmonics by their complex counterpart using (\ref{real_complex}).  
If we expand in terms of complex spherical harmonics, we find
\begin{equation}\label{melonii}
\mathlarger{\mathlarger{\ominus}}=    \frac{1}{3\pi^2} \times  (2\pi)^3b^2\sum_{\bold{l}\geq \textbf{2},\bold{m}} \frac{\mathcal{W}_{\bold{l},\bold{m}} \mathcal{W}^*_{\bold{l},\bold{m}}}{ \, A_{l_1} A_{l_2} A_{l_3}}~,
\end{equation}
where $\bold{l} = (l_1,l_2,l_3)$, $\bold{m} = (m_1,m_2,m_3)$, and
\begin{eqnarray}
\mathcal{W}_{\bold{l},\bold{m}} &\equiv&  \int\dd\Omega \, \mathcal{Y}_{l_1 m_1}(\Omega)\mathcal{Y}_{l_2 m_2}(\Omega)\mathcal{Y}_{l_3 m_3}(\Omega) \\ 
&=& \sqrt{\frac{(2l_1+1)(2l_2+1)(2l_3+1)}{4\pi}} 
\begin{pmatrix}
l_1 & l_2 & l_3\\
0 & 0 & 0
\end{pmatrix}
\begin{pmatrix}
l_1 & l_2 & l_3\\
m_1 & m_2 & m_3
\end{pmatrix}~, 
\end{eqnarray}
where in the second line we have written it in terms of Wigner 3-j symbols. $\mathcal{W}^*_{\bold{l},\bold{m}}$ is the complex conjugate. Combining the orthogonality relation of the 3j symbol with the fact that each of the $m_i'$s itself run over $2l_i+1$ integers we obtain
\begin{equation}\label{m}
\mathlarger{\mathlarger{\ominus}} = \frac{2}{3} b^2\sum_{\bold{l}\geq \textbf{2}} \frac{(2l_1+1)(2l_2+1)(2l_3+1)}{A_{l_1}A_{l_2} A_{l_3}} \begin{pmatrix} l_1 & l_2 & l_3 \\ 0 & 0 & 0\end{pmatrix}^2~.
 \end{equation}~
\newline
Now we combine all the diagrams. In particular we realise that all UV divergences cancel out, i.e.
\begin{equation}
\bigcirc\!\!\bigcirc_1+ \bigcirc\!\!\bigcirc_2+ \bigcirc\!\!\bigcirc_3 =0~,\quad \bigcirc_1+\bigcirc_2=0~.
\end{equation}
We are left with the finite result
\begin{multline}\label{eq:2loop}
\frac{1}{\mathcal{Z}^{(2)}_{\text{pert}}[b]}\times \mathcal{Z}_{\text{pert}}[b] = 1+ \frac{2}{3} b^2\sum_{\bold{l}\geq \textbf{2}} \frac{(2l_1+1)(2l_2+1)(2l_3+1)}{A_{l_1}A_{l_2} A_{l_3}} \begin{pmatrix} l_1 & l_2 & l_3 \\ 0 & 0 & 0\end{pmatrix}^2-\frac{1}{2}b^2 \sum_{l\geq 2}\frac{(2l+1)(2l+3)}{A_l^2}\cr
+\frac{1}{6}b^2\sum_{l\geq 2}\frac{(2l+1)^2}{A_l^2}= 1+ \left( \mathlarger{\mathlarger{\ominus}} + \bigcirc\!\bigcirc\right)b^2+ \ldots~.
\end{multline}
We can evaluate the two cactus sums explicitly yielding at leading order for $A_l= (l(l+1)-2)$
\begin{equation}
\bigcirc\!\!\bigcirc=-\frac{1}{2}\sum_{l\geq 2}\frac{(2l+1)(2l+3)}{A_l^2}+ \frac{1}{6}\sum_{l\geq 2}\frac{(2l+1)^2}{A_l^2} = -\frac{11}{27}-\frac{\pi^2}{9}~.
\end{equation}
\subsection{Final two-loop result}
Collecting all the results, in particular combining the saddle (\ref{Zsaddle}), the one-loop contribution (\ref{eq:1loop_semi}) and the two-loop contribution (\ref{eq:2loop}) we obtain
\begin{multline}\label{finalZgrav}
\mathcal{Z}^{(0)}_{\text{grav}}[\upsilon]= \pm e^{2\vartheta}\times \mathcal{A}\times \mathrm{const}\times \upsilon^{-\frac{1}{b^2}-2}\,\Lambda_{\text{uv}}^{-\frac{1}{b^2}-1}e^{-4\pi \Lambda\upsilon}\cr
\times \left(\frac{1}{b^4}+\left(\frac{7}{6}+ \frac{5a_1}{2a_0}\pi +2\gamma_E+ \log(4\pi)+ \mathlarger{\mathlarger{\ominus}} + \bigcirc\!\bigcirc\right)\frac{1}{b^2}+ \ldots \right)~,
\end{multline}
where the constant we defined in (\ref{const}) and $\mathcal{A}$ we defined below (\ref{g0Z}). Before comparing the above semiclassical expansion of the sphere partition function of two-dimensional quantum gravity at fixed area obtained from a path integral with the sphere partition function obtained from the three point function of three area operators $\mathcal{O}_b\equiv e^{2b\varphi}$ we discuss the individual pieces of (\ref{finalZgrav}). 
\begin{itemize}
\item The power of $\upsilon$ contains the one-loop contribution; the $l=0,1$ and $2$ modes which we treated separately and the $\upsilon$ dependency of the functional determinant. The area constraint contributes an inverse power of $\upsilon$. Finally, both the ghost and matter sphere partition functions, constituting two-dimensional conformal field theories contribute $(\upsilon\Lambda_{\text{uv}})^{(c_\text{m}-26)/6}$. 
\item Up to two-loop the Feynman diagrams do not contribute any divergences. We expect this to hold true also for higher order Feynman diagrams. 
\end{itemize}

\section{Sphere partition function from DOZZ}\label{sec:DOZZ}
In this section we compare our result (\ref{finalZgrav}) with the sphere partition function obtained from the three-point function of three area operators $\mathcal{O}_b= e^{2b\varphi}$
\begin{equation}
\langle \mathcal{O}_{b}(z_1) \mathcal{O}_{b}(z_2) \mathcal{O}_{b}(z_3) \rangle =  \frac{1}{\text{vol}_{PSL(2,\mathbb{C})}} \times \frac{\mathcal{C}(b,b,b)}{|z_1-z_2|^2 |z_1-z_3|^2 |z_2-z_3|^2}~,
\end{equation}
where \cite{Zamolodchikov:1995aa,Dorn:1994xn,Teschner:2001rv,Teschner:1995yf} 
\begin{equation}\label{dozz}
\mathcal{C}(b,b,b) 
= -\Lambda^{Q/b-3}(\pi\gamma(b^2))^{Q/b}\frac{(1-b^2)^2}{\pi^3b^5 \gamma(b^2)\gamma(b^{-2})}e^{-Q^2+ Q^2\log 4}~.
\end{equation}
In this expression $\gamma(z) \equiv \Gamma(z)/\Gamma(1-z)$ is a meromorphic function with poles when $z$ is a non-positive integer and zeros when $z$ is a non-vanishing positive integer. The sphere partition function follows upon integrating the DOZZ coefficient three-times with respect to the cosmological constant and setting the integration constants to zero\footnote{The integration constants are analytic expressions in $\Lambda$. One way to understand this ambiguity would be through comparison with the matrix integral sphere partition function. Apart from the critical exponents, the matrix free-energy contains ambiguous analytic pieces.}
\begin{equation}
- \partial_\Lambda^3\mathcal{Z}_{\mathrm{DOZZ}}[\Lambda]= 2 \times \mathcal{C}(b,b,b)~,
\end{equation}
where we have used that \cite{Zamolodchikov:1995aa}
\begin{equation}
\int_{\mathbb{C}^3} \frac{\dd^2 z_1 \dd^2 z_2 \dd^2 z_3}{|z_1-z_2|^2 |z_1-z_3|^2 |z_2-z_3|^2} =  2 \, {\text{vol}_{PSL(2,\mathbb{C})}}~.
\end{equation}
We obtain
\begin{equation}\label{Z_DOZZ}
\mathcal{Z}_{\mathrm{DOZZ}}[\Lambda]= 2\times \left(\pi \Lambda \gamma(b^2)\right)^{Q/b}\frac{(1-b^2)}{\pi^3 Q \gamma(b^2)\gamma(b^{-2})}e^{-Q^2 +Q^2 \log 4}~.
\end{equation}
This expression is similar to the expressions encountered e.g. in \cite{Zamolodchikov:1995aa,Dorn:1994xn}; the difference being the factor $e^{-Q^2 +Q^2 \log 4}$ which appears when transforming the Liouville action from the sphere to the disk using \cite{Harlow:2011ny}
\begin{equation}
\varphi \rightarrow \varphi- Q \log\frac{2}{1+z\bar{z}}~.
\end{equation}
On the disk of radius $\kappa\gg 1$ we obtain 
\begin{equation}
S_L^{(\mathrm{Disk})}[\varphi]= \frac{1}{4\pi}\int_D \dd r \dd \theta \sqrt{\tilde{g}_D} \left(\tilde{g}^{ab}_D \partial_a\varphi \partial_b \varphi + 4 \pi \Lambda\, e^{2b\varphi}\right)+ \frac{Q}{\pi}\oint_{\partial D}\dd \theta\varphi+ 2Q^2 \log \kappa + Q^2 - Q^2\log 4~,
\end{equation}
where $d \tilde{s}^2_D= \dd r^2 + r^2\dd\theta^2$ is the flat metric on the disk $D$. 

To compare (\ref{Z_DOZZ}) with the path integral result  (\ref{finalZgrav})  we restore units. Since Liouville theory itself without fixing the area is a two-dimensional CFT the dependency of the partition function on the area $\upsilon_0$ of the sphere is fixed to equal $\upsilon_0^{c_L/6}$. By dimensional analysis we obtain
\begin{equation}\label{eq:ZDOZZ}
\mathcal{Z}_{\mathrm{DOZZ}}[\Lambda]=2\upsilon_0^{\frac{c_L}{6}}\Lambda_{\text{uv}}^{\frac{7}{6}+b^2}\,\left(\pi \Lambda \gamma(b^2)\right)^{Q/b}\frac{(1-b^2)}{\pi^3 Q \gamma(b^2)\gamma(b^{-2})}e^{-Q^2 +Q^2 \log 4}~.
\end{equation}
\subsection{Fixed area $\&$ comparison}
To compare with  (\ref{finalZgrav}), stemming from (\ref{g0Z}) we complexify the cosmological constant $\Lambda \rightarrow \Lambda - i\alpha$, and calculate \cite{Zamolodchikov:2005jb}
\begin{multline}
\mathcal{Z}_{\mathrm{DOZZ, \,fix}}[\upsilon]=\int_{\mathbb{R}} \frac{\dd\alpha}{2\pi}\, e^{-4\pi i\upsilon \alpha}\mathcal{Z}_{\mathrm{DOZZ}}[\Lambda- i\alpha]\cr
=2\upsilon_0^{\frac{c_L}{6}}\Lambda_{\text{uv}}^{\frac{7}{6}+b^2}\, e^{-4\pi \Lambda\upsilon}\left(\frac{\gamma(b^2)}{4}\right)^{Q/b}\frac{\Gamma(2-b^2)}{4\pi^4b^3 \Gamma(b^2)\Gamma(b^{-2})}e^{-Q^2 +Q^2 \log 4}\,\upsilon^{-1-Q/b}~.
\end{multline}
In the semiclassical $b\rightarrow 0$ limit we obtain
\begin{equation}
\mathcal{Z}_{\mathrm{DOZZ, \,fix}}[\upsilon]\approx \frac{\sqrt{2}}{\pi^{9/2}}\,e^{-2-2\gamma}\,\upsilon_0^{\frac{c_L}{6}}\Lambda_{\text{uv}}^{\frac{7}{6}+b^2}\, \,e^{-4\pi \Lambda\upsilon}\,\upsilon^{-\frac{1}{b^2}-2}\left(\frac{1}{b^4}+ \left(-\frac{25}{12}+ 2\log 2\right)\frac{1}{b^2}+ \ldots \right)~.
\end{equation}
We can now compare the fixed area transformed DOZZ sphere partition function (\ref{Z_DOZZ}) with the path integral expression (\ref{finalZgrav}). Since in particular in the DOZZ formula the UV length scale has been set to one and we only re-instored it in (\ref{eq:ZDOZZ}) through dimensional analysis, to compare with the path integral picture we should allow for a scheme dependent length scale. We will assume $\Lambda_{\text{uv}}= s\tilde{\Lambda}_{\text{uv}}$, leading to 
\begin{eqnarray}\label{comparison}
\mathcal{Z}_{\mathrm{DOZZ, \,fix}}[\upsilon]&\approx&  \frac{\sqrt{2}}{\pi^{9/2}}\,s^{\frac{7}{6}}\,e^{-2-2\gamma}\,\upsilon_0^{\frac{c_L}{6}}\tilde{\Lambda}_{\text{uv}}^{\frac{7}{6}+b^2}\, \,e^{-4\pi \Lambda\upsilon}\,\upsilon^{-\frac{1}{b^2}-2}\left(\frac{1}{b^4}+ \left(-\frac{25}{12}+ 2\log 2+\log s\right)\frac{1}{b^2}+ \ldots \right)~\cr
\mathcal{Z}^{(0)}_{\text{grav}}[\upsilon]&\approx& \pm e^{2\vartheta}\times \mathcal{A}\times \mathrm{const}\times \upsilon^{-\frac{1}{b^2}-2}\,e^{-4\pi \Lambda\upsilon}\Lambda_{\text{uv}}^{-\frac{1}{b^2}-1}\cr
&\times& \left(\frac{1}{b^4}+\left(\frac{7}{6}+ \frac{5a_1}{2a_0}\pi +2\gamma_E+ \log(4\pi)+  \mathlarger{\mathlarger{\ominus}}+ \bigcirc\!\bigcirc\right)\frac{1}{b^2}+ \ldots \right)~.
\end{eqnarray}
Consequently the two expressions agree upon setting 
\begin{equation}\label{s}
s= \frac{1}{\pi}\,e^{\frac{13}{4}+\frac{5a_1}{2a_0}\pi +2\gamma_E+ \ominus+ \bigcirc\!\bigcirc}~.
\end{equation}
We now compare the two expressions (\ref{comparison})
\begin{itemize}
\item
 From (\ref{comparison}) we infer that the Fadeev-Popov determinant has an important effect on the semiclassical expansion of the sphere partition function. One might therefore wonder whether our gauge choice is permissible. Following the ideas of \cite{Distler:1988jt} we believe that since we perform a semiclassical analysis one can avoid the Gribov ambiguities \cite{Gribov:1977wm}. In particular since in a semiclassical expansion the Fadeev-Popov determinant (\ref{FPexact}) would only change sign if the Gaussian fluctuations compete at order $\mathcal{O}(Q)$, our semiclassical expansion $Q\rightarrow \infty$ seems provides a loophole. It would be interesting to explore whether our gauge choice breaks down at lower orders in the semiclassical expansion. 
\item
Furthermore we note that we do not encounter any sign ambiguity in the semiclassical expansion of the DOZZ formula. This combined with the explicit calculation performed in \ref{app:unsuppressed} suggests that out of the two possible Wick rotations (\ref{Wickrotation}) only case a) which Wick rotates the two unsuppressed Gaussian modes (\ref{interactions2})  oppositely is consistent. It would be interesting to see if we can find further evidence of this at higher order in the loop expansion. 
\item We fix (\ref{s}) at two-loop order. It would be interesting to include higher order contributions to test (\ref{s}).
\end{itemize}
Consequently, including also the contribution of the ghost and matter CFT,  we conjecture the two-sphere partition function for two-dimensional quantum gravity at fixed area
\begin{equation}\label{eq:Zgrav}
\mathcal{Z}_{\text{grav}}^{(0)}[\upsilon] = e^{2\vartheta}\times 2\mathcal{A}\times \upsilon^{-\frac{1}{b^2}-2} \Lambda_{\text{uv}}^{-\frac{1}{b^2}-1}\, e^{-4\pi \Lambda\upsilon}\left(\frac{\gamma(b^2)}{4}\right)^{\frac{Q}{b}}\frac{\Gamma(2-b^2)}{4\pi^4b^3 \Gamma(b^2)\Gamma(b^{-2})}e^{-Q^2 +Q^2 \log 4}\,~.
\end{equation}
\subsection{Comparison case A) $\&$ case B)}
We finish this section by comparing the two-sphere partition functions in the two possible cases mentioned in the introduction (sec.\ref{sec:intro}). Whereas two-dimensional quantum gravity is a priori a heavily fluctuating set-up it allows for two possibilities to tame the fluctuations.
Case A) we studied in this paper leading to (\ref{eq:Zgrav}), case B) was studied in \cite{timelike}. We now compare the two results\footnote{We note that the factor of $2$ is a normalisation convention. Additionally we introduce the UV length-scale in case B) to highlight the fact that it is a dimensionless quantity.}
\begin{eqnarray}\label{conj}
\mathcal{Z}_{\text{grav},A}^{(0)}[\upsilon] &=& e^{2\vartheta}\times 2\mathcal{A}\times \upsilon^{-\frac{1}{b^2}-2} \Lambda_{\text{uv}}^{-\frac{1}{b^2}-1}\, e^{-4\pi \Lambda\upsilon}\left(\frac{\gamma(b^2)}{4}\right)^{\frac{Q}{b}}\frac{\Gamma(2-b^2)}{4\pi^4b^3 \Gamma(b^2)\Gamma(b^{-2})}e^{-Q^2 +Q^2 \log 4}~, \cr
\mathcal{Z}^{(0)}_{\text{grav},B}[\Lambda]  &=&\mp   e^{2\vartheta}\times \mathcal{A} \times \Lambda_{\text{uv}}^{\frac{1}{\beta^2}-1}\left(\pi \Lambda \gamma(-\beta^2)\right)^{-\frac{1}{\beta^2} +1}\frac{(1+\beta^2)}{q \gamma(-\beta^2)\gamma(-\beta^{-2})}\,e^{q^2 -q^2 \log 4}~,
\end{eqnarray}
where $(b,Q)$ are related to $(\beta,q)$ by $(b,Q) \rightarrow (\pm i\beta, \mp iq)$. Even before expanding in the semiclassical limit we note the sign ambiguity in (\ref{conj}). Whereas for the fixed area case A) we find two unsuppressed Gaussians (\ref{interactions2}) whose Jacobians cancel each, in the timelike case only the $l=0$ mode leads to an unsuppressed Gaussian. The possibility to Wick rotate this by $\pm\pi/2$ up or down is encoded in the sign in $\mathcal{Z}^{(0)}_{\text{grav},B}[\Lambda]$. The semiclassical (small $b$ and small $\beta$ respectively) expansion reveals further differences; we have 
\begin{eqnarray}\label{st}
\mathcal{Z}_{\text{grav},A}^{(0)}[\upsilon] &\approx& e^{2\vartheta}\times \mathcal{A}\times \upsilon^{-\frac{1}{b^2}-2}\Lambda_{\text{uv}}^{-\frac{1}{b^2}-1}\, e^{-4\pi \Lambda\upsilon}\left(\frac{1}{b^4}+ \ldots \right)~,\cr
\mathcal{Z}^{(0)}_{\text{grav},B}[\Lambda]  &\approx &\pm i e^{2\vartheta}\times \mathcal{A} \times \Lambda_{\text{uv}}^{\frac{1}{\beta^2}-1}\Lambda^{-\frac{1}{\beta^2}+1}\,\left( 1 - e^{\frac{2i\pi}{\beta^2}}\right)\times e^{-\frac{1}{\beta^2}- \frac{1}{\beta^2}\log (4\pi \beta^2)}\,\left(\frac{1}{\beta}+ \ldots \right)~.
\end{eqnarray}
In both cases the Fadeev-Popov determinant contributes a cubic inverse power of $b$ or $\beta$ respectively. In case A) the rescaling of $\delta\alpha$ (\ref{Spert}) contributes another power, yielding the final $\mathcal{O}(1/b^4)$. In case B) on the other side the saddle point itself contributes a quadratic power in $\beta$, yielding the expansion $\mathcal{O}(1/\beta)$. The details are explained in \cite{timelike}. Furthermore from (\ref{st}) we infer that in case B) two saddles contribute to the partition function---one of them heavily oscillating and in particular leading to a vanishing sphere partition function for positive integer values of $1/\beta^2$. In the spacelike case on the other side, only one real saddle (\ref{saddle}) contributes to the fixed area two-sphere partition function $\mathcal{Z}_{\text{grav},A}^{(0)}[\upsilon] $.

\section{Outlook $\&$ matrix integrals}
Our expression (\ref{finalZgrav}) provides the genus zero semiclassical expansion of the two-sphere partition function of two-dimensional gravity at fixed area. Following the hypothesis of Distler-Kawai \cite{Distler:1988jt} and David \cite{David:1988hj} for arbitrary genus $h$ we obtain 
\begin{equation}\label{Zh}
\log \mathcal{Z}^{(h)}_{\text{grav}}[\upsilon]= \chi_h \vartheta- \frac{\chi_h}{24}\left(\sqrt{(c_\text{m}-25)(c_\text{m}-1)}-c_\text{m}+25\right)\log\upsilon + f^{(h)}(c_\text{m})~,
\end{equation}
where we absorbed $\upsilon^{-1}e^{-4\pi \upsilon\Lambda}$ in the measure. For large and negative $c_\text{m}$ we obtain
\begin{equation}
\log \mathcal{Z}^{(h)}_{\text{grav}}[\upsilon]\approx \chi_h \vartheta+ \frac{\chi_h}{2}\left(\frac{c_{\text{m}}}{6}- \frac{19}{6}+ \ldots \right)\log\upsilon + f^{(h)}(c_\text{m})~,
\end{equation}
In particular, even though by adding the fixed area constraint we broke conformality on a genus zero surface the gravitational two-sphere partition function scales as $\upsilon^{c_\text{m}/6}$ \cite{Anninos:2020ccj,Anninos:2020geh}, reminiscent of the entanglement entropy of a two-dimensional CFT  \cite{Holzhey:1994we, Calabrese:2004eu, Casini:2011kv}. The subleading terms could then be viewed as the effects of coupled the CFT to gravity. 
Furthermore using
\begin{equation}
c_\text{m}= -\frac{6}{b^2}+13-6b^2~,
\end{equation}
we reproduce the semiclassical expansion  (\ref{finalZgrav}). As mentioned in the introduction two-dimensional quantum gravity coupled to the series of minimal models $\mathcal{M}_{2,2m-1}$ is conjectured to be dual to multicritical matrix integrals \cite{Kazakov:1989bc,Staudacher:1989fy}. A multicritical matrix integral consists of a Hermitian $N\times N$ matrix ordered in an even order $2m$ potential with $(m-1)$ real valued couplings
\begin{equation}
V_{m}(M)=\frac{1}{2}M^2+ \frac{1}{4}\alpha_2 M^4+ \ldots + \frac{1}{2m}\alpha_m M^{2m}~.
\end{equation}
One then studies the large $N$ limit of the matrix integral
\begin{equation}
\int_{\mathbb{R}^{N^2}}[\mathcal{D}M]\, e^{-N^2\mathrm{Tr}_{N\times N}V_{\text{m}}(M)}~.
\end{equation}
We highlight that the multicritical matrix integral has $(m-1)$ real couplings, which is equal to the number of primaries of $\mathcal{M}_{2,2m-1}$. 
In particular in \cite{Anninos:2020geh, db} we constructed the path coupling space leading to the identity operator, reproducing the coefficient of the logarithm in (\ref{Zh}). 
The main object of observation in this paper was $f^{(0)}(c_\text{m})$. If the conjectured duality between matrix integrals and two-dimensional quantum gravity coupled to $\mathcal{M}_{2,2m-1}$ holds true it should be possible to also match these expressions with the matrix integral. Clearly in that case restricting to a single genus is not enough as one will encounter ambiguities in the normalisation. However upon taking ratios of different genera, ambiguities can be avoided.

\section*{Acknowledegements}
It is a pleasure to acknowledge Teresa Bautista and Lorenz Eberhardt for useful discussions. I would particularly like to thank Dio Anninos for infinite discussions and his endless support and encouragement  $\href{https://arxiv.org/abs/2004.01171}{\tikz\penguin[hat=magenta, scale=0.15];}$. This work would not have been possible without it. The work of B.M. is part of the research programme of the Foundation for Fundamental Research on Matter (FOM), which is financially supported by the Netherlands Organisation for Science Research (NWO). 

\appendix

\section{Spherical harmonics}\label{Yapp}
We use real valued spherical harmonics throughout the paper. We denote by $\mathcal{Y}_{l m}(\theta,\phi)$ the complex spherical harmonics defined by 
\begin{equation}
\mathcal{Y}_{l m}(\theta,\phi)= \sqrt{\frac{(2l+1)}{4\pi}\frac{(l-m)!}{(l+m)!}}\, P_{l,m}(\cos\theta)\, e^{im\phi}~,
\end{equation}
where $P_{l,m}$ is the associated Legendre function, and $m \in [-l,l]$ with $l\in\mathbb{N}$. Real spherical harmonics $Y_{l m}(\theta,\phi)$ can be obtained using the linear combinations
\begin{equation}\label{real_complex}
{Y}_{l m}(\theta,\phi)= \begin{cases}
\frac{i}{\sqrt{2}}\left(\mathcal{Y}_{l m}(\theta,\phi)- (-1)^m \mathcal{Y}_{l, -m}(\theta,\phi)\right)~,\quad \mathrm{if}~ m<0\\
\mathcal{Y}_{l 0}(\theta,\phi)\\
\frac{1}{\sqrt{2}}\left(\mathcal{Y}_{l, -m}(\theta,\phi)+ (-1)^m \mathcal{Y}_{l m}(\theta,\phi)\right)~,\quad \mathrm{if}~ m>0~.
\end{cases}
\end{equation}
The Wigner 3-j symbol is given by the Clebsch-Gordan coefficients and gives the integral of the product of three complex spherical harmonics
\begin{align}
&\int \dd\phi \dd\theta\sin\theta\,\mathcal{Y}_{l_1,m_1}(\theta,\phi)\mathcal{Y}_{l_2,m_2}(\theta,\phi)\mathcal{Y}_{l_3,m_3}(\theta,\phi)\cr
&= \sqrt{\frac{(2l_1+1)(2l_2+1)(2l_3+1)}{4\pi}}\begin{pmatrix} l_1 & l_2 & l_3 \\ 0 & 0 & 0\end{pmatrix}\begin{pmatrix} l_1 & l_2 & l_3 \\ m_1 & m_2 & m_3\end{pmatrix}~.
\end{align}
\textbf{3-j symbol relations.}
The Clebsch-Gordan coefficients satisfy various properties. In particular they obey the orthogonality relation
\begin{equation}\label{orthoCG}
\sum_{\alpha,\beta}\begin{pmatrix} a & b & c \\ \alpha & \beta & \gamma\end{pmatrix}\begin{pmatrix} a & b & c' \\ \alpha & \beta & \gamma'\end{pmatrix}= \frac{1}{2c+1}\delta_{cc'}\delta_{\gamma\gamma'}~.
\end{equation}
Furthermore
\begin{equation}\label{reduced_CG}
\begin{pmatrix}
a & b & c \\ 
0 & 0 & 0 
\end{pmatrix} \neq 0\quad  \mathrm{iff} ~a+ b+ c \in 2\mathbb{Z}~\quad \&\quad \begin{pmatrix}
a & b & 0 \\ 
\alpha & \beta & 0 
\end{pmatrix}= \frac{(-1)^{a-\alpha}}{\sqrt{2a+1}}\delta_{ab}\delta_{\alpha-\beta}~.
\end{equation}

\section{Unsuppressed Gaussians}\label{app:unsuppressed}
Instead of performing Wick rotations we evaluate (\ref{Zpert}) keeping the unsuppressed $l=0$. 
We have 
\begin{multline}\label{unsuppressed}
\mathcal{Z}_{\text{pert}}^{(2)}[b]=\int [\mathcal{D}'\delta\varphi]\,e^{-\hat{S}^{(2)}_{\text{pert}}[\delta\varphi]}\times e^{\frac{1}{2\pi}\delta\varphi_{00}^2}\,\int_{\mathbb{R}} \frac{\dd \delta\alpha}{2\pi} \, e^{{4ib\sqrt{\pi}\delta\alpha}\delta\varphi_{00}}\times e^{\frac{1}{4\pi}\int\dd\Omega\,\left(\frac{4}{3}b\delta\varphi(\Omega)^3+\frac{2}{3}b^2\delta\varphi(\Omega)^4\ldots \right)}\cr
\times \sum_{n=0}^\infty \frac{(i\delta\alpha)^n \left(\int\dd{\Omega}(e^{2b\delta\varphi(\Omega)}-2b\delta\varphi(\Omega))-4\pi\right)^n}{n!}~,
\end{multline}
where $S_{\text{pert}}^{(2)}[\delta\varphi]$ is the Gaussian action containing all but the $l=0$ and $l=1$ modes, additionally keeping track of the contribution of the Fadeev-Popov determinant to the $l=2$ modes. 
We can now shift $\delta \alpha$
\begin{equation}
\delta \alpha \rightarrow 4{b}{\sqrt{\pi}} \delta\alpha~.
\end{equation}
Applying 
 \begin{equation} \int_{\mathbb{R}} \dd x\, e^{ix y}(ix)^n= (\partial_y)^n \delta(y)\end{equation}
to (\ref{unsuppressed}) and integrating $n$ times by parts (dropping the boundary terms) we obtain
\allowdisplaybreaks
\begin{multline}\label{eq:setup_int}
\mathcal{Z}_{\text{pert}}^{(2)}[b]=\frac{1}{8\pi^{3/2} b}\int [\mathcal{D}'\delta\varphi]\,e^{-S^{(2)}_{\text{pert}}[\delta\varphi]}\,\times \sum_{n=0}^\infty\,\frac{(-1)^n}{n!} \left(\frac{1}{ 4{b}{\sqrt{\pi}}}\right)^n \,\cr
\times  \partial_{\delta\varphi_{00}}^n\, \left[{\left(\int\dd{\Omega} (2b^2\delta\varphi(\Omega)^2 + \frac{4}{3}b^3\delta\varphi(\Omega)^3+\ldots  ) \right)^n}\times e^{\frac{1}{4\pi}\int\dd\Omega\,\left(\frac{4}{3}b\delta\varphi(\Omega)^3+\frac{2}{3}b^2\delta\varphi(\Omega)^4\ldots \right)}\times e^{\frac{1}{2\pi}\delta\varphi_{00}^2}\right]_{\delta\varphi_{00}=0}~.
\end{multline}
Evaluating the above for $n=0,1$ and $n=2$ we reproduce (\ref{eq:2loop}).

 \begingroup


\addcontentsline{toc}{section}{References}
\section*{References}
\begin{thebibliography}{99}


  
\bibitem{Distler:1988jt} 
  J.~Distler and H.~Kawai,
  ``Conformal Field Theory and 2D Quantum Gravity,''
  Nucl.\ Phys.\ B {\bf 321}, 509 (1989).
  doi:10.1016/0550-3213(89)90354-4
  
\bibitem{David:1988hj} 
  F.~David,
  ``Conformal Field Theories Coupled to 2D Gravity in the Conformal Gauge,''
  Mod.\ Phys.\ Lett.\ A {\bf 3}, 1651 (1988).
  doi:10.1142/S0217732388001975
  

\bibitem{Polyakov:1981rd}
A.~M.~Polyakov,
``Quantum Geometry of Bosonic Strings,''
Phys. Lett. B \textbf{103}, 207-210 (1981)
doi:10.1016/0370-2693(81)90743-7

\bibitem{Zamolodvarphikov:1982vx} 
  A.~B.~Zamolodchikov,
  ``On The Entropy Of Random Surfaces,''
  Phys.\ Lett.\  {\bf 117B}, 87 (1982).
  doi:10.1016/0370-2693(82)90879-6


\bibitem{db}

D. Anninos and B.~M\"uhlmann,
``Matrix integrals \& the two-sphere,''
to appear

\bibitem{dS_musings}
D.~Anninos,
``De Sitter Musings,''
Int. J. Mod. Phys. A \textbf{27}, 1230013 (2012)
doi:10.1142/S0217751X1230013X
[arXiv:1205.3855 [hep-th]].


\bibitem{Polyakov:1984yq}
A.~M.~Polyakov, A.~A.~Belavin and A.~B.~Zamolodchikov,
``Infinite Conformal Symmetry of Critical Fluctuations in Two-Dimensions,''
J. Statist. Phys. \textbf{34}, 763 (1984)
doi:10.1007/BF01009438
  
\bibitem{Friedan}
D.~Friedan, Z.~a.~Qiu and S.~H.~Shenker,
``Conformal Invariance, Unitarity and Two-Dimensional Critical Exponents,''
Phys. Rev. Lett. \textbf{52}, 1575-1578 (1984)
doi:10.1103/PhysRevLett.52.1575

\bibitem{Seiberg:1990eb}
N.~Seiberg,
``Notes on quantum Liouville theory and quantum gravity,''
Prog. Theor. Phys. Suppl. \textbf{102}, 319-349 (1990)
doi:10.1143/PTPS.102.319

\bibitem{Zamolodchikov:2005fy}
A.~B.~Zamolodchikov,
``Three-point function in the minimal Liouville gravity,''
Theor. Math. Phys. \textbf{142}, 183-196 (2005)
doi:10.1007/s11232-005-0003-3
[arXiv:hep-th/0505063 [hep-th]].

   
\bibitem{Zamolodchikov:1995aa} 
  A.~B.~Zamolodchikov and A.~B.~Zamolodchikov,
  ``Structure constants and conformal bootstrap in Liouville field theory,''
  Nucl.\ Phys.\ B {\bf 477}, 577 (1996)
  doi:10.1016/0550-3213(96)00351-3
  [hep-th/9506136].


\bibitem{Dorn:1994xn} 
  H.~Dorn and H.~J.~Otto,
  ``Two and three point functions in Liouville theory,''
  Nucl.\ Phys.\ B {\bf 429}, 375 (1994)
  doi:10.1016/0550-3213(94)00352-1
  [hep-th/9403141].
 
\bibitem{Teschner:2001rv} 
  J.~Teschner,
  ``Liouville theory revisited,''
  Class.\ Quant.\ Grav.\  {\bf 18}, R153 (2001)
  doi:10.1088/0264-9381/18/23/201
  [hep-th/0104158].
  
\bibitem{Teschner:1995yf}
J.~Teschner,
``On the Liouville three point function,''
Phys. Lett. B \textbf{363}, 65-70 (1995)
doi:10.1016/0370-2693(95)01200-A
[arXiv:hep-th/9507109 [hep-th]].




\bibitem{timelike}
D.~Anninos, T.~Bautista and B.~M\"uhlmann,
``The two-sphere partition function in two-dimensional quantum gravity,''
[arXiv:2106.01665 [hep-th]].
 
 
 \bibitem{Kazakov:1989bc}
V.~Kazakov,
``The Appearance of Matter Fields from Quantum Fluctuations of 2D Gravity,''
Mod. Phys. Lett. A \textbf{4}, 2125 (1989)
doi:10.1142/S0217732389002392

\bibitem{Staudacher:1989fy}
M.~Staudacher,
``The Yang-lee Edge Singularity on a Dynamical Planar Random Surface,''
Nucl. Phys. B \textbf{336}, 349 (1990)
doi:10.1016/0550-3213(90)90432-D
 
 
\bibitem{Anninos:2020geh}
D.~Anninos and B.~M\"uhlmann,
``Matrix integrals $\&$ finite holography,''
[arXiv:2012.05224 [hep-th]].

\bibitem{Anninos:2020ccj}
D.~Anninos and B.~M\"uhlmann,
``Notes on matrix models (matrix musings),''
J. Stat. Mech. \textbf{2008}, 083109 (2020)
doi:10.1088/1742-5468/aba499
[arXiv:2004.01171 [hep-th]].
  
\bibitem{Polchinski:1989fn}
J.~Polchinski,
``A Two-Dimensional Model for Quantum Gravity,''
Nucl. Phys. B \textbf{324}, 123-140 (1989)
doi:10.1016/0550-3213(89)90184-3



  



\bibitem{Zamolodchikov:2001dz}
A.~Zamolodchikov,
``Scaling Lee-Yang model on a sphere. 1. Partition function,''
JHEP \textbf{07}, 029 (2002)
doi:10.1088/1126-6708/2002/07/029
[arXiv:hep-th/0109078 [hep-th]].


\bibitem{loooongSphere}
D.~Anninos, F.~Denef, Y.~T.~A.~Law and Z.~Sun,
``Quantum de Sitter horizon entropy from quasicanonical bulk, edge, sphere and topological string partition functions,''
[arXiv:2009.12464 [hep-th]].

\bibitem{Gibbons:1976ue}
G.~W.~Gibbons and S.~W.~Hawking,
``Action Integrals and Partition Functions in Quantum Gravity,''
Phys. Rev. D \textbf{15}, 2752-2756 (1977)
doi:10.1103/PhysRevD.15.2752
  
\bibitem{Gibbons:1977mu}
G.~W.~Gibbons and S.~W.~Hawking,
``Cosmological Event Horizons, Thermodynamics, and Particle Creation,''
Phys. Rev. D \textbf{15}, 2738-2751 (1977)
doi:10.1103/PhysRevD.15.2738



\bibitem{Adamchik}
Victor S. Adamchik, Polygamma functions of negative order,
Journal of Computational and Applied Mathematics,
Volume 100, Issue 2,
1998, Pages 191-199,
ISSN 0377-0427, https://doi.org/10.1016/S0377-0427(98)00192-7


\bibitem{Holzhey:1994we} 
  C.~Holzhey, F.~Larsen and F.~Wilczek,
  ``Geometric and renormalized entropy in conformal field theory,''
  Nucl.\ Phys.\ B {\bf 424}, 443 (1994)
  doi:10.1016/0550-3213(94)90402-2
  [hep-th/9403108].
  
\bibitem{Calabrese:2004eu} 
  P.~Calabrese and J.~L.~Cardy,
  ``Entanglement entropy and quantum field theory,''
  J.\ Stat.\ Mech.\  {\bf 0406}, P06002 (2004)
  doi:10.1088/1742-5468/2004/06/P06002
  [hep-th/0405152].
  
\bibitem{Casini:2011kv} 
  H.~Casini, M.~Huerta and R.~C.~Myers,
  ``Towards a derivation of holographic entanglement entropy,''
  JHEP {\bf 1105}, 036 (2011)
  doi:10.1007/JHEP05(2011)036
  [arXiv:1102.0440 [hep-th]].
 
  
\bibitem{Gribov:1977wm}
V.~N.~Gribov,
``Quantization of Nonabelian Gauge Theories,''
Nucl. Phys. B \textbf{139}, 1 (1978)
doi:10.1016/0550-3213(78)90175-X

\bibitem{Harlow:2011ny}
D.~Harlow, J.~Maltz and E.~Witten,
``Analytic Continuation of Liouville Theory,''
JHEP \textbf{12}, 071 (2011)
doi:10.1007/JHEP12(2011)071
[arXiv:1108.4417 [hep-th]].

\bibitem{Gibbons:1978ac}
G.~W.~Gibbons, S.~W.~Hawking and M.~J.~Perry,
``Path Integrals and the Indefiniteness of the Gravitational Action,''
Nucl. Phys. B \textbf{138}, 141-150 (1978)
doi:10.1016/0550-3213(78)90161-X

\bibitem{Polchinski:1988ua}
J.~Polchinski,
``The Phase of the Sum Over Spheres,''
Phys. Lett. B \textbf{219}, 251-257 (1989)
doi:10.1016/0370-2693(89)90387-0



\bibitem{Bros:2010wa}
J.~Bros, H.~Epstein and U.~Moschella,
``Scalar tachyons in the de Sitter universe,''
Lett. Math. Phys. \textbf{93}, 203-211 (2010)
doi:10.1007/s11005-010-0406-4
[arXiv:1003.1396 [hep-th]].


\bibitem{Vassilevich:2003xt}
D.~V.~Vassilevich,
``Heat kernel expansion: User's manual,''
Phys. Rept. \textbf{388}, 279-360 (2003)
doi:10.1016/j.physrep.2003.09.002
[arXiv:hep-th/0306138 [hep-th]].



\bibitem{Zamolodchikov:2005jb}
A.~B.~Zamolodchikov,
``Perturbed conformal field theory on fluctuating sphere,''
[arXiv:hep-th/0508044 [hep-th]].


 

\bibitem{Zamolodchikov:2001ah}
A.~B.~Zamolodchikov and A.~B.~Zamolodchikov,
``Liouville field theory on a pseudosphere,''
[arXiv:hep-th/0101152 [hep-th]].





 
%


%











%
%
%
%
%
%
%

%
%
%

  
  
\end{thebibliography}
\end{document}